\documentclass[twocolumn,aps,ams,epsfig,graphicx,showpacs,floatfix,amssymb,prl]{revtex4}
\usepackage{graphicx}
\newcommand{\be}{\begin{equation}}
\newcommand{\ee}{\end{equation}}

\newcommand{\bea}{\begin{eqnarray}}
\newcommand{\eea}{\end{eqnarray}}
\newcommand{\bd}{\begin{displaymath}}
\newcommand{\ed}{\end{displaymath}}
\newcommand{\bi}{\begin{itemize}}
\newcommand{\ei}{\end{itemize}}
\newcommand{\bc}{\begin{center}}
\newcommand{\ec}{\end{center}}
\newcommand{\bfl}{\begin{flushleft}}
\newcommand{\efl}{\end{flushleft}}
\newcommand{\bfr}{\begin{flushright}}
\newcommand{\efr}{\end{flushright}}
\newcommand{\f}{\frac}

\def\br{{\bf r}}\def\bR{{\bf R}}
\def\bk{{\bf k}} \def\bK{{\bf K}} \def\bp{{\bf p}}

\def\6{\partial}  
 \def\d{\delta} \def\ve{\varepsilon}

  \def\D{\Delta}

\def\={\!\!\!&=&\!\!\!}
\def\+{\!\!\!&&\!\!\!+~}
\def\-{\!\!\!&&\!\!\!-~}

\begin{document}

\author{I. \c{T}ifrea$^{1}$, M. Poggio$^{2,3}$, D.D. Awschalom$^3$, and M. E. Flatt\'{e}$^4$}

\affiliation{$^1$Department of Physics and Astronomy, California State University, Fullerton, CA 92834, USA}
\affiliation{$^2$IBM Research Division, Almaden Research Center, San Jose CA, 95120, USA\\ and Center for Probing the Nanoscale, Stanford University, Stanford CA, 94305, USA}
\affiliation{$^3$Center for Spintronics and Quantum Computing,
University of California, Santa Barbara, CA 93106, USA}
\affiliation{$^4$Department of Physics and Astronomy, University of Iowa, Iowa City, IA
52242, USA}
\date{\today}
\title{Overhauser frequency shifts in semiconductor nanostructures}

\begin{abstract}
 We calculate the Overhauser frequency shifts in semiconductor nanostructures resulting from the hyperfine interaction between   nonequilibrium electronic spins and nuclear spins. The frequency shifts depend on the electronic local density of states and spin   polarization as well as the electronic and nuclear spin relaxation mechanisms. Unlike previous calculations, our method accounts for the electron confinement in low dimensional semiconductor nanostructures, resulting in both nuclear spin polarizations and Overhauser shifts that are strongly dependent on position. Our results explain previously puzzling measurements of Overhauser shifts in an Al$_x$Ga$_{1-x}$As parabolic quantum well by   showing the connection between the electron spin lifetime and the frequency shifts.
\end{abstract}
%
\maketitle


Nuclear spins in low dimensional semiconductor nanostructures have
been identified as a potential basis for spin-based electronic devices
\cite{wolf} or quantum information processing
\cite{vandersypen}. Structures such as quantum wells or quantum dots
provide a natural environment for the efficient manipulation of
nuclear spin dynamics \cite{martinoAPL,bracker}. The control is
indirect and relies on the hyperfine interaction between nuclear spins
in the crystalline lattice and electronic spins in the conduction
band. The efficiency of the hyperfine interaction depends on the
overlap between electronic and nuclear  wave functions, such that
control over the electronic population can be reflected in the
dynamics of the nuclear spin population \cite{overhauser}.

In the presence of magnetic fields and polarized optical excitation,
both the electronic and nuclear spin systems become polarized. The
optically induced polarization of the electronic spin system is
transferred to the nuclei via the hyperfine interaction --- a phenomenon known
as dynamical nuclear polarization (DNP) --- leading to an enhanced
nuclear spin polarization. This nonequilibrium nuclear polarization
can be detected as either an Overhauser or Knight shift in the
electron or nuclear Larmor frequencies respectively. DNP effects are
typically detectable by electron spin resonance or nuclear magnetic
resonance (NMR) in bulk systems where a large number of nuclei is present \cite{slichter}.

Detection of the nonequiliubrium nuclear spin polarization in semiconductor nanostructures is exceptionally challenging due to the very small number of nuclei involved.
Optical techniques developed first for bulk systems \cite{paget} can
be used to overcome this problem \cite{barrett}. Optical pumping of
the electronic population leads to an enhanced nuclear spin
polarization which can improve the resolution of NMR detection.
Minimum detection volumes can be reduced from volumes containing
10$^{17}$ nuclear spins, required by standard NMR techniques, to
ensembles comprised by as few as 10$^5$ nuclear spins, enough to
resolve single quantum wells or quantum dots
\cite{martinoAPL,bracker}. The particular geometry of these systems
also provides the basis for DNP manipulation \cite{tifreaJS,tifreaNP}.
In certain geometries, electrostatic control over spin-polarized
electrons can be used to selectively polarize regions of nuclear spin
\cite{tifrea,martino}.

The efficiency of such methods was shown in time-resolved Faraday rotation (TRFR) experiments performed in an Al$_x$Ga$_{1-x}$As parabolic quantum well (PQW). When subject to external electric fields, the electron density can be localized at different positions along the confinement direction of the well.  As a result, DNP can be made to occur in different parts of the well depending on the applied electric field. TRFR experiments in a 100-nm Al$_x$Ga$_{1-x}$As PQW demonstrated the ability to imprint a nuclear polarization profile at the position of the polarized electron density. Depending on the applied gate voltage, the resulting 8-nm wide distribution of nuclear polarization could be centered at different positions over a 20-nm range within the PQW \cite{martino}.

Here we describe the theoretical basis of Overhauser shifts in systems with reduced dimensionality, where electron confinement cannot be ignored.  We show our approach is essential to a successful quantitative calculation of the spin precession frequencies of electrons subject to an induced position-dependent nuclear spin polarization by comparing the calculations to measurements on the 100-nm PQW described above \cite{martino}. The central quantities
determining the induced nuclear spin polarization are the initial
electronic spin polarization from optical pumping, the electronic spin
relaxation rate, the position dependent nuclear spin relaxation time
due to the hyperfine interaction, and the additional nuclear spin
relaxation time due to mechanisms other than the hyperfine
interaction. The induced nuclear spin polarization is responsible for
both hyperfine and dipolar magnetic fields, which along with the
applied external magnetic field determine the value of the electronic
spin precession frequency. All quantities involved, except the
optically induced electronic spin polarization, are position dependent
and can be controlled with external electric fields. We find that the
dependence of the frequency shifts on the electron spin lifetime, and
the electric field dependence of the electron spin lifetimes, explain
the qualitative trends of the frequency shifts observed in
Ref.~\onlinecite{martino}, as well as their magnitudes within a factor
of 2.

DNP as a result of the hyperfine interaction between electronic
and nuclear spins in semiconductor quantum wells and quantum dots
produces a position dependent nuclear spin
polarization\cite{overhauser,tifreaJS}:
\be\label{NP}
{\cal P}_I(\br_n)=\f{\sum_m m e^{m\mu_{ind}(\br_n)}}{I\sum_m
e^{m\mu_{ind}(\br_n)}}\;,
\ee
where $I$ is the nuclear spin quantum number, $m=I,I-1,\cdots,-I$, and
\bea\label{mu}
\mu_{ind}(\br_n)&=&\f{1}{k_BT\int d\ve A^2_e(\br_n,\ve)
f'_{FD}(\ve)}\nonumber\\&&\times
\f{T'_n}{T_{1n}(\br_n)+T'_n}\f{D_0-D}{N_n}\;.
\eea
Here, $k_B$ is the Boltzmann constant, $T$ the temperature, $A_e(\br,\ve)=\sum_l|\psi_m(\br)|^2\d(\ve-E_l)$ is the electronic local density of states at the nuclear position $\br_n$ ($l$ labels the state, $E_l$ its energy, and $\psi_l(\br)$ is the electronic wave function), $f_{FD}(\ve)$ the Fermi-Dirac function, $T_{1n}(\br_n)$ the nuclear spin relaxation time due to the hyperfine interaction, and $T'_n$ the additional nuclear spin relaxation time due to mechanisms other than the hyperfine interaction, $D=N_+-N_-$ is the electron spin polarization with $D_0$ being its thermal equilibrium value for the applied external magnetic field, and $N_n$ is the nuclear density. The induced nuclear spin polarization was obtained assuming a time independent electronic spin polarization in the system \cite{tifreaJS}. In turn, DNP, and implicitly the induced nuclear spin polarization, affects the properties of the electron spin system. Most notably, this nuclear spin polarization results in an Overhauser shift of the electronic spin precession frequency, $\D\nu^e_{hf}$, which can be calculated  as
\be\label{OS}
\D\nu^e_{hf}(\br)=-\f{2\mu_0}{3 h} g_0 \mu_B\mu_n\sum_i
g_n^i|\psi^i(\br)|^2c_i(\br)<{\bf I}_i(\br)>\;,
\ee
where $\mu_0$ is the vacuum permeability, $h$ the Planck constant,
$g_0$ the bare electronic g-factor, $\mu_B$ the electronic Bohr
magneton, and $\mu_n$ the nuclear Bohr magneton. The summation runs
over all different nuclear species characterized by the nuclear
g-factor $g_n^i$, a position dependent concentration $c_i(\br)$, and
an average nuclear spin polarization $<{\bf I}_i(\br)> \cong {\cal
  P}^i_I(\br)$. The total electronic wave function, $\psi^i(\br)$  at
the position of {\bf the} nuclei will also depend on the nuclear species. Dipole nuclear spin interactions will induce an additional shift in the electronic spin precession frequencies \cite{tifreaNP}. Such shifts are usually much smaller than the Overhauser shifts and will be neglected hereafter. Unlike the hyperfine shifts, the dipolar shifts are proportional to the effective electronic g-factor, and they may be important in semiconductor systems characterized by large values of the electronic g-factor.

Eq. (\ref{OS}) states that in samples with reduced dimensionality the Overhauser shift in the electronic spin precession frequency is position dependent. In general, frequency shifts measured in TRFR constitute a response of the entire electronic system, and therefore their value depends on the probability of finding electrons at various positions in the sample. Accordingly, we expect that data from TRFR experiments will represent a convolution of the position dependent Overhauser shift with the square of the electronic wave function,
\be\label{OSExp}
\D\nu_L=\f{\int d\br \D\nu^e_{hf}(\br) |\psi(\br)|^2 }{\int d\br
|\psi(\br)|^2 }\;.
\ee
\begin{figure}[t]
\centering \scalebox{0.75}[0.75]{\includegraphics*{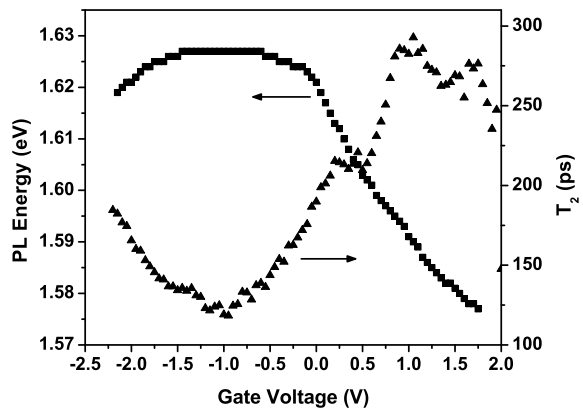}}
\caption{Integrated PL energy ($\blacksquare$) and the transverse
    electron spin life time $T_2$ ($\blacktriangle$) as function of
    applied gate voltage in a 100-nm PQW.}
\label{fig1}
\end{figure}

In fact, experiments support this expectation. Frequency shifts are measured by TRFR in an undoped 100-nm (100) Al$_x$Ga$_{1-x}$As PQW grown by molecular beam epitaxy \cite{kato}. The aluminum concentration $x$ is varied from 7\% at the center of the well to 40\% in the barriers to create a parabolic confinement potential in the conduction band \cite{miller}. A voltage $U_g$ applied across the front gate and the grounded back gate produces a constant electric field across the PQW and results in a negligible leakage current ($< 100 \mu$A). Fig. \ref{fig1} shows the emission energy of photoluminescence (PL) as a function of $U_g$ at T = 5 K excited by a pulsed laser at 1.676 eV with and average intensity of 1 W/cm$^2$. The PL energy is highest at the built in potential $U_0$ = -1.1 $\pm$ 0.1 V and decreases as $U_g$ is tuned away from that value. Such behavior is consistent with the quantum-confined Stark effect \cite{dab1} and shows that the application of $U_g$ affects the band-structure as expected. The observed shift is 6.8 $\pm$1.0 meV/V$^2$ compared with a theoretical value of 5.0 meV/V$^2$ using conduction and valance band offsets from the literature \cite{dab2} and neglecting changes in the exciton binding energy, which are of the order of 4 meV \cite{chuliang}.

TRFR measurements are performed using a 76 MHz femtosecond Ti:Sapphire laser tuned near the absorption edge of the PQW (1.62 eV at $U_g = 0$ V). The measurement, which monitors small rotations in the linear polarization of the probe beam transmitted through a sample, is sensitive to the direction of spin polarization of electrons in the conduction band.  Electron spin precession is well described by the expression for TRFR,
\be\label{fitFR}
\Theta_F(\D t)=\Theta_\perp e^{-\D t/T^*_2}\cos{(2\pi\nu_L\D
t+\phi)+\Theta_\parallel e^{-\D t/T_{1e}}}\;,
\ee
where $\Theta_\perp$ is proportional to the spin injected perpendicular to the applied field, $\Theta_\parallel$ is proportional to the spin injected parallel to the applied field, $T_2^*$ is the inhomogeneous transverse spin lifetime, and $T_{1e}$ is the longitudinal spin lifetime. The Larmor frequency $\nu_L = g \mu_B B/h + \Delta \nu_L$ depends on the applied field $B$ and the Overhauser shift $\Delta \nu_L$ due to nuclear polarization.  Fits to TRFR data using Eq. (\ref{fitFR}) result in the extraction of a number of parameters including $T_2^*$ and $\Delta \nu_L$. $T_2^*$ is plotted in Fig. 1a as a function of $U_g$ and reaches a minimum near $U_0$ reflecting the formation of excitons in the flat-band condition due to the spatial proximity of the confined electrons and holes. The reduction of $T_2^*$ can be attributed to the enhanced charge recombination rate of bound electrons and holes or to electron-hole exchange which may also limit the electron spin lifetime at low temperature \cite{paillard}. $T_{1e}$, whose dependence on $U_g$ closely follows $T_2^*$, plays an important role in determining the nuclear polarization $<I>$ in our samples, since in the dynamic nuclear polarization (DNP) process $<I>\propto<S_\parallel>$, the average electron spin polarization along $B$. Since $<S_\parallel>\propto T_{1e}$, larger overall nuclear polarizations are achieved for $U_g$ far from rather than close to $U_0$. In the experiment nuclei were polarized at constant gate voltages $U_g$ (0 V, 0.5 V, and 1 V) corresponding to different positions of the maximum initial nuclear spin polarization across the PQW (3 nm, 7 nm, and 10 nm from the quantum well's center). After the induced nuclear polarization was saturated (about 20 min), TRFR data were collected quickly enough not to disturb the built-up polarization. By making measurements at different $U_g$, $\nu_L$ is measured  with the electron population overlapping different regions of the induced nuclear spin polarization profile. The result is the measurement of an Overhauser shift which depends on $U_g$, i.e., a position dependent Overhauser frequency shift across the PQW \cite{martino}.
\begin{figure}[t]
\centering \scalebox{0.8}[0.8]{\includegraphics*{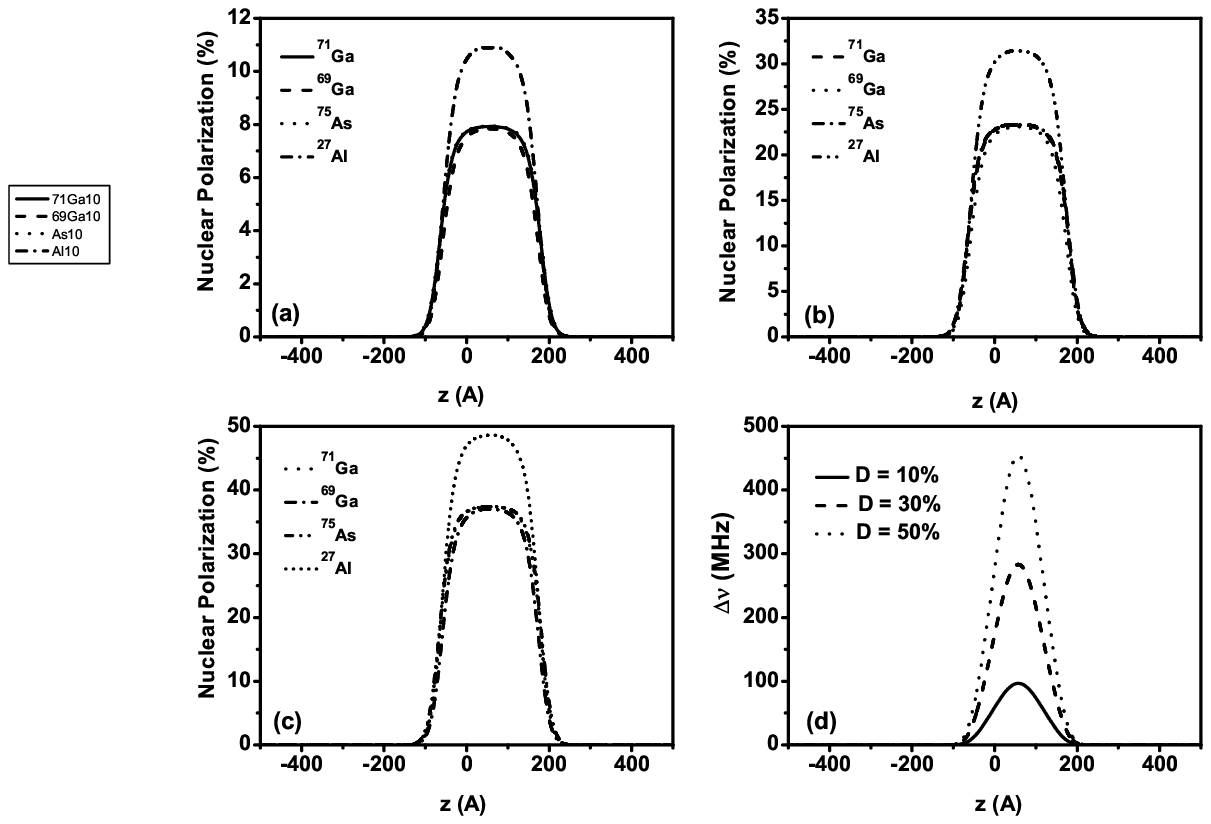}}
\caption{The position dependence of the induced nuclear spin polarization in the Al$_x$Ga$_{1-x}$As PQW for $^{71}$Ga, $^{69}$Ga, $^{75}$As, and  $^{27}$Al nuclei calculated for different values of the initial electronic spin polarization (a)$D=10\%$; b)$D=30\%$, and c)$D=50\%$), and d) the corresponding total Overhauser frequency shifts.}
\label{fig2}
\end{figure}

The dispersion relations for electrons in a PQW are
quasi-two dimensional and the total electronic wave function can be written as a product between an envelope function, $\phi(z)$, and a Bloch function, $u(\br)$,
\be\label{elfunction}
\psi_{j\bK}(\br_n)=\exp{[i\bK\cdot\bR]\;\phi_j(z)\;u(\br_n)}\;.
\ee
Based on this assumption the electronic local density of states is
\be\label{QWELDOS}
A_e(\br_n,\ve)=\sum_{j} |\phi_{j}({\bf z}_n)|^2\;N_{2D}\Theta(\ve - E_{j(\bK=0)})\;,
\ee
where $N_{2D}$ is the density of states for a two-dimensional electron gas and $\Theta(z)$ is the Heavyside step function. The value of the electronic envelope function is evaluated using a numerical $\bk\cdot\bp$ calculation \cite{wayne}. The Bloch function takes different values for different nuclear species. For both $^{71}$Ga and $^{69}$Ga nuclei its value was evaluated to be $|u(\br_n)|^2=5.2\times10^{25}$ cm$^{-3}$ \cite{tifrea}, not very different from the value reported in bulk systems \cite{paget}. For $^{75}$As nuclei we will consider the Bloch function calculated in bulk systems, namely $|u(\br_n)|^2=9.8\times10^{25}$ cm$^{-3}$. $^{27}$Al nuclei replace in the PQW both $^{71}$Ga and $^{69}$Ga and we will assume the corresponding Bloch function at the aluminum sites to be the same as for gallium nuclei. For all numerical calculations we consider only the first electronic conduction subband to be occupied. The additional nuclear spin relaxation time due to mechanism other than the hyperfine interaction is of the order of $T_n' \sim 600$ s and is considered to be temperature independent \cite{mcneil}. Our calculation accounts also for the internal, sample related, voltage observed in PL experiments, $U_0$.

Figure \ref{fig2} presents the calculated induced nuclear spin polarization in the Al$_x$Ga$_{1-x}$As PQW used in the experiments for different values of the initial electronic spin polarization. The reported nuclear spin polarization is obtained in the saturation regime of the DNP process \cite{tifreaNP}. We consider the gate voltage to be 0 V, so the displacement of the nuclear spin polarization from the PQW center is attributed to the internal, growth related voltage. The identified active nuclear species are $^{71}$Ga ($I=3/2$), $^{69}$Ga ($I=3/2$), $^{75}$As ($I=3/2$), and $^{27}$Al ($I=5/2$). For an initial electronic spin polarization of 50\%, nuclear species characterized by a nuclear quantum number $I=3/2$ will develop a maximum nuclear spin polarization of the order of 40\%. On the other hand,  $^{27}$Al nuclei, characterized by $I=5/2$, will present a larger maximum nuclear spin polarization of about 50\%.  The width of the region containing polarized nuclei is about 30 nm and is related to the value of the additional nuclear spin relaxation time $T_n'$. The calculated induced nuclear spin polarization will increase with the efficiency of the optical pumping process, namely with the electron spin polarization.
\begin{figure}[t]
\centering \scalebox{1}[1]{\includegraphics*{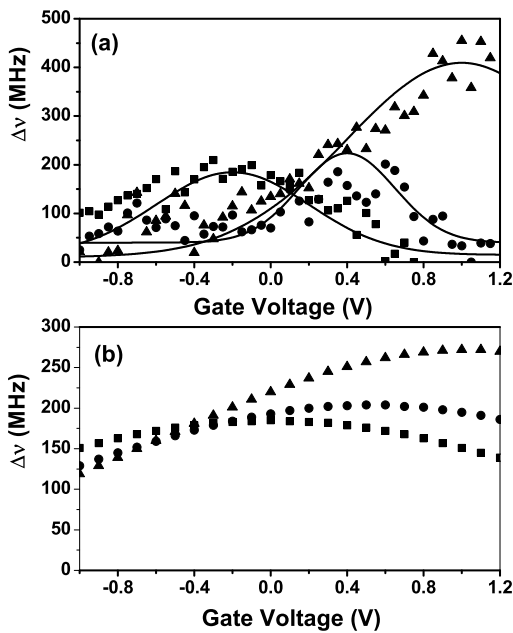}}
\caption{a) Measurements of Overhauser frequency shifts for
    different initial nuclear spin polarizations ($\blacksquare$ - 0
    V, $\bullet$ - 0.5 V, $\blacktriangle$ - 1 V) - lines are included
    for eye guidance; b) Calculations of Overhauser frequency shifts
    for an initial $\sim$28\% electronic spin polarization
    ($\blacksquare$ - 0 V, $\bullet$ - 0.5 V, $\blacktriangle$ - 1
    V).}
\label{fig3}
\end{figure}

Each of the active nuclear species contributes to the total Overhauser shift of the electron spin precession frequency. The particular geometry of the sample makes this contribution for Al and Ga nuclei position-dependent, namely $c_{Al}=1.14\times(-1.08 + [1.31+2.99592\cdot 10^{-6} z^2]^{1/2})$, with $z$ the position across the PQW measured in \AA, and $c_{Ga}=1-c_{Al}$. For Ga nuclei the different isotopes, $^{71}$Ga and $^{69}$Ga, will be characterized by different natural abundances, namely 39.89\% and 60.1\%. Figure \ref{fig3}a presents the experimental Overhauser shifts observed in TRFR measurements with nuclei initially polarized at $U_g =$ 0, 0.5, and 1.0 V.  As we already mentioned, these frequency shifts should be viewed as a response to the polarization of the whole sample due to the overlap of the electron probability density with the nuclear polarization profile within the PQW. Figure \ref{fig3}b presents the calculated Overhauser shift as a function of $U_g$ for an initial electron spin polarization of about 28\% and same initial conditions for the nuclei in the sample. The choice of 28\% for the optically pumped electronic spin polarization is made to most closely match the experimental data for $U_g=0$ V and is reasonable given the experimental conditions.  This polarization induces a maximum of about 22\% nuclear spin polarization for Ga and As nuclei, and about 30\% nuclear spin polarization for Al nuclei, corresponding to an average nuclear spin polarization of about 2.5\% across the PQW. The induced nuclear spin polarization will also be gate voltage dependent as the pumped  electron spin polarization changes according to the value of  the electronic spin relaxation time $T_{1e}$. The theoretical calculation neglects nuclear spin diffusion effects, which for semiconductor nanostructures are minimal \cite{martino}. The calculations of frequency
shifts agree well with the TRFR data presented in Figure \ref{fig3}a.

In conclusion, we present a theoretical investigation of the
Overhauser frequency shifts in samples with reduced dimensionality.
Our calculations for a gated 100-nm PQW predict a position-dependent
nuclear polarization whose position in the confinement direction can
be determined by the external gate voltage. We calculated the
resulting position-dependent Overhauser shift which agrees well with
experimental data obtained by TRFR measurements. This agreement
between experiment and theory is consistent with our ability to use
localized spin-polarized electrons to orient nanometer-scale profiles
of nuclear spin. Extending such techniques from systems with one-dimensional confinement (quantum wells) to systems with higher degrees of confinement (quantum wires,  quantum dots) may allow for the controllable polarization of drastically fewer nuclear spins.

I.T. would like to acknowledge the support of Research Corporation. M.E.F. and D.D.A. would like to acknowledge the support of an ONR MURI. M.P. and D.D.A. would like to acknowledge also the support of NSF.

\end{document}